\documentstyle[12pt,preprint,eqsecnum,aps,tighten]{revtex}

\def\be{\begin{equation}}
\def\ee{\end{equation}}
\def\ba{\begin{array}{c}}
\def\ea{\end{array}}

\def\ben{$$}

\def\een{$$}

\begin{document}

\title{Anomalous doublets of states \\
in a
${\cal PT}$ symmetric quantum model}

\author{Miloslav Znojil}

\address{Nuclear Physics Institute of Academy of Sciences of the Czech
Republic, 250 68 \v Re\v z, Czech Republic}

\author{G\'eza L\'evai}

\address{Institute of Nuclear Research of the Hungarian Academy of
Sciences, PO Box 51, H--4001 Debrecen, Hungary}

\author{Pinaki Roy and Rajkumar Roychoudhury}

\address{Physics and Applied Mathematics Unit, Indian Statistical
Institute, \\ Calcutta 700035, India }

\maketitle

\begin{abstract}
We complexify one of the Natanzon's exactly solvable potentials in
${\cal PT}$ symmetric manner and discover that it supports the
pairs of bound states with the same number of nodal zeros. This
could indicate that the Sturm Liouville oscillation theorem does
not admit an immediate generalization.

\end{abstract}

\vspace{1cm}


\newpage

\section{Framework: ${\cal PT}$ symmetric quantum mechanics}
\label{odin}

Bound states in a smooth {\em real} potential $V(x)\geq 0$ are
most easily interpreted in the language of Sturm--Liouville theory
\cite{Sturm}. Its oscillation theorems imply that the $N-$th bound
state $\psi_N(x)$ possesses $N$ nodal zeros. Even the standard
boundary conditions may be understood as the presence of an {\em
additional} pair of zeros which are located at both ends of the
interval of coordinates $ {\cal J}= (-\infty,\infty)$ in one
dimension or of the radial axis $ {\cal J}= (0,\infty)$ in three
dimensions.

A non-standard, additional constraint is only necessary for some
strongly singular forces \cite{Landau}. Unfortunately, the latter,
mathematically rigorous requirement of the elimination of the
irregular solutions may prove fairly counterintuitive and, for
this reason, it is often being forgotten in practice. For
illustration, we may recollect the paper on supersymmetry
\cite{Fred} which describes a {singular} map of the shifted
harmonic oscillator. This example (attributed to A. Khare in
Acknowledgements) is marred by a subtle violation of the boundary
condition in the origin. A very similar inconsistency appears in
the ``conditionally exact" model by A. de Souza Dutra
\cite{Dutra}, in some phenomenological studies in quantum
chemistry \cite{japrvni} etc.

An extensive clarification of the latter point may be found
elsewhere~\cite{ja}. In order to avoid similar misunderstandings,
the underlying mathematics can be significantly simplified via and
analytic continuation (or even re-formulation) of the physical
boundary conditions \cite{Turbiner}. This is an innovative idea
which proved unexpectedly fruitful. Recently, one of its more
specific versions became a cornerstone of the so called ${\cal
PT}$ symmetric approach to quantum mechanics \cite{BBjmp}.

The latter formalism dispenses with the Hermiticity of the
Hamiltonian $H$ and keeps only a weaker requirement of the
commutativity of $H$ with the product of parity (${\cal P} x =
-x$) and complex conjugation (${\cal T} i = -i$ which mimics the
time reversal). The singular Schr\"{o}dinger equations may be then
regularized in the way which preserves the reality of the
spectrum. Such a ${\cal PT}$ symmetric non-Hermitization can be
applied easily to many solvable potentials, e.g., via a constant
complex shift of the coordinate axis \cite{Gezaz},
 \be
r = r_{(t)} = t - i\,\varepsilon, \ \ \ \ \ \ t \in (-\infty,
\infty), \ \ \ \ \ \varepsilon > 0.
 \label{shift}
 \ee
This enables us, in essence,

\begin{itemize}

\item
to work, more easily, with the analytic wave functions $\psi(r)$
which prove available within the whole complex plane of $r$,

\item
to return, whenever necessary, to the Hermitian Hamiltonian by
means of the limiting transition $\varepsilon \to 0$ and/or of a
suitable selection of the additional constraints.

\end{itemize}

 \noindent
Both the one-dimensional and three-dimensional exactly solvable
harmonic oscillators $V^{(E)}(r)= \ell(\ell+1)/r^2+r^2$ with
parity $\ell=-1,0$ or angular momentum $\ell = 0, 1, \ldots$,
respectively, become suddenly tractable on equal footing (details
may be found in ref. \cite{ptho}). In the three-body context of
the so called Calogero's exactly solvable model \cite{Calogero},
the same singular harmonic-oscillator-like differential equation
and the same limiting transition $\varepsilon \to 0$ reproduce the
correct Hermitian solutions in spite of the utterly different
physical origin and meaning of the singular term \cite{Tater}. Via
a modified transition to the Hermitian limit, another solvable
model is revealed~\cite{solvable}.

One can summarize that the ${\cal PT}$ symmetric quantum mechanics
offers a new approach to the explicit construction of bound
states. It will be applied here to the potential
 \be
 V^{}(\xi)=
  \frac{3/4}{(1-e^{2i\xi})^2}
 +\frac{2\,i\,\beta\,\exp(2i\xi)}{\sqrt{1-e^{2i\xi}}}
 -\frac{C}{1-e^{2i\xi}}
 \label{hulth}
 \ee
with certain unusual features. Already Singh and Devi \cite{Singh}
have noticed that it does not belong to the family of the shape
invariant potentials. Dutt et al \cite{Dutt} emphasized that this
is the only existing nontrivial example of the conditionally
exactly solvable potential. We have recently shown \cite{newRR}
that this exactly solvable potential of the Natanzon class
\cite{Natanzon} possesses a natural supersymmetric interpretation
and offers a nontrivial opportunity of the rigorous study of its
solutions near the singularity. The latter point proved in fact
also the main motivation of our continuing interest in this
example.


\section{The method: A change of coordinates}
\label{dvah}

In a way which dates back to the work of Liouville
\cite{Liouville} one can change variables in the Schr\"{o}dinger
equation
 \be
\left[-\,\frac{d^2}{dr^2} + V^{(E)}(r)\right]\, \psi^{(E)}(r) =
E^{(E)}\,\psi^{(E)}(r)
 \label{SEor}
 \ee
using the recipe
 \be
 r \to \xi=\xi(r), \ \ \ \ \ \ \ \ \ \
 \psi^{(E)}(r) \to
\psi^{(D)}(\xi)
 = \sqrt{\xi'[r(\xi)]}\cdot \psi^{(E)}[r(\xi)].
 \ee
This generates the new differential equation of the similar form,
 \be
\left[-\,\frac{d^2}{d\xi^2} + V^{(D)}(\xi)\right]\,
\psi^{(D)}(\xi) = E^{(D)}\,\psi^{(D)}(\xi)\, .
 \label{SEoxi}
 \ee
The explicit relationship between the two respective potentials is
determined by the identity which is easily derived,
 \be
V^{(E)}(r)- E^{(E)}=
 \left [
 \xi'(r)
 \right ]^2
 \left \{
V^{(D)}(\xi) -E^{(D)}
 \right \}
+
 \frac{3}{4}
 \left [
 {\xi''(r) \over \xi'(r)}
 \right ]^2
 -
 \frac{1}{2}
 \left [
 {\xi'''(r) \over \xi'(r)}
 \right ].
 \label{newpot}
  \ee
Once you postulate the solvability of the original equation
(\ref{SEor}) in terms of some known polynomials, it is possible to
construct another exactly solvable equation by the suitable choice
of the re-parametrization $\xi(r)$. Its special cases which
mediate all the mutual canonical transformations between the {\em
real} shape invariant potentials were listed in review
\cite{Khare} (cf., in particular, Figure 5.1 there).

Once we switch our attention to the non-Hermitian examples and
retain the split of the shape invariant models into the
Laguerre-solvable and Jacobi-solvable subsets (cf. Figure 5.1 of
ref. \cite{Khare} once more), we can easily parallel many of the
above transformations. For the Laguerre-solvable subset, all the
details may be found in refs. \cite{Morse} and \cite{Coulomb}. In
a less exhaustive manner, the Jacobi-solvable shape invariant
${\cal PT}$ symmetric family has been described in refs.
\cite{Gezaz} and \cite{Eckart}. One of the most striking features
of some of the complexified changes of variables lies in the
characteristic bent shape of the integration contours $\xi$. For
illustration, let us recollect the implicit definition
 \be
 \sinh r= - i e^{i\xi}
\label{tren}
 \ee
of the Hulth\'{e}n-force generating function $\xi=\xi(r)$ as
employed in ref. \cite{Eckart}. Its straight-line input
$r=r_{(t)}= t - i\,\varepsilon$ with $\varepsilon>0$ becomes
{strongly} deformed after the transition to
$\xi=\xi(r)=\xi(r_{(t)})$.

In more detail, the left path could also inessentially be
deformed, for the sake of simplicity, in such a way that
$\varepsilon=\varepsilon(t) \to 0$ for $|t| \to \infty$. This
asymptotically modified contour of $r=r_{(t)}$ avoids the
upwards-running cut and comfortably coincides with the real line
at both its asymptotic ends. It remains parametrized by $t \in
(-\infty,\infty)$ and defines the right-hand side bent curve $\xi
= \Omega - iZ$ via the pair of the real implicit equations
 \ben
 \ba
 \sinh t \cos \varepsilon(t) = e^Z\,\sin \Omega,\\
 \cosh t \sin \varepsilon(t) = e^Z\,\cos \Omega.
 \ea
 \een
Their explicit solution
 \ben
 \ba
  \Omega_{(t)} =\arctan \left [
\frac{\tanh t}{\tan \varepsilon(t)} \right ] \in \left
(\Omega_{(-\infty)}, \Omega_{(\infty)}\right )\equiv \left
(-\frac{\pi}{2}+\varepsilon(-\infty),
\frac{\pi}{2}-\varepsilon(\infty) \right ),\\  Z_{(t)} =
\frac{1}{2} \ln \left [ \sinh^2t+\sin^2\varepsilon(t)
 \right ]
 \ea
 \een
shows that the new curve $\xi =\xi(t)=\Omega_{(t)}-iZ_{(t)}$ is
arch-shaped and symmetric. It starts in a left imaginary minus
infinity (at $Z_{(-\infty)} \gg 1$) and ends in its right parallel
(with $\Omega_{(\infty)} =\pi/2-\varepsilon(\infty)$ and
$Z_{(\infty)} \gg 1$). Its top $Z_{(0)}= \ln \sin \varepsilon <0$
at $t=\Omega_{(0)}=0$ moves upwards in an inverse proportion to
the decrease of the original shift $\varepsilon(0) \to 0$.


\section{Pairs of states with the same $N$} \label{trih}

In a search for a ``new" solvable model $V^{(D)}(\xi)$, let us
start from the complexified Schr\"{o}dinger equation (\ref{SEor})
defined on a complex contour $r=r_{(t)}= t - i\,\varepsilon(t)$.
We shall apply the change of the complex contours (\ref{tren}) to
the complexified Eckart potential
 \be
 V^{(E)}(r) = \frac{A(A-1)}{\sinh^2 r}-2\,i\,\beta\frac{\cosh
r}{\sinh r}, \ \ \ \ \ \ \ \ \ \ \beta>0 .
 \label{Eckartp}
 \ee
Although our result will coincide (not surprisingly) with the
above-mentioned shape-non-invariant potential (\ref{hulth}), the
mapping itself will be shown to exhibit certain very unusual and
unexpected features.

In a preparatory step, the exact solvability of such a model may
be most easily demonstrated via an auxiliary re-parametrization
suggested in ref. \cite{Eckart},
 \ben
  \psi^{(E)}
  (r) = (y-1)^u (y+1)^v \varphi \left (\frac{1-y}{2} \right ),
  \ \ \ \ \ \ \ \ \
y=y(r)=\frac{\cosh r}{\sinh r}=1-2z
  \een
with
 \ben
 4v^2=2i\beta-E, \ \ \ \ \  \ \ \
4u^2=-2i\beta-E.
  \een
This leads to the new differential equation
\be
z(1-z)\,\varphi''(z) +[c-(a+b+1)z]\,\varphi'(z) -ab\,\varphi(z)=0
 \ee
which is completely solvable in terms of the Gauss hypergeometric
series,
 \be
\varphi(z)= C_1\cdot\ _2F_1(a,b;c;z) + C_2\cdot\ z^{1-c}\
 _2F_1(a+1-c,b+1-c;2-c;z)
 \label{solu}
  \ee
where
 \ben
 c=1+2u, \ \ \ \ a+b=2u+2v+1, \ \ \ \
ab=(u+v)(u+v+1)+A(1-A).
  \een
All the freedom of parameters becomes fixed by the boundary
conditions $\psi(\pm \infty) =0$ which give $b=-N$ and  $a = b \pm
(2A-1)$. Jacobi polynomials $P^{(\alpha,\beta)}(x)$ \cite{Abram}
enter the elementary formula for the wave functions,
 \be \psi^{(E)}(r) =
 \left ( \frac{1}{\sinh r} \right )^{u+v} \ e^{(v-u)r}
  \cdot P^{(u/2,v/2)}_N(\coth r).
 \label{soluc1}
  \ee
These solutions remain normalizable if and only if
 \be
 a=2A-N-1, \ \ \ \ u+v=A-N-1>0, \ \ \ \
 u-v=-i\,\frac{\beta}{A-N-1}.
 \label{proto}
 \ee
The model possesses $N_{max}< A-1$ bound states with the energies
 \be
 E^{(E)} =
 -\left ( A-N-1 \right )^2 + \frac{\beta^2}{(A-N-1)^2},
 \ \ \ \ \ \ N = 0, 1, \ldots, N_{max}.
 \label{spectr}
 \ee
We are prepared to transform this solution into its partner of eq.
(\ref{SEoxi}). It suffices to change variables via eq.
(\ref{tren}). This replaces the Eckart problem (\ref{SEor}) by the
new Schr\"{o}dinger equation (\ref{SEoxi}). The implicit
definition (\ref{newpot}) of the new potential $V^{(D)}(\xi)$
acquires a more explicit form
 \ben
 V^{(D)}(r)- E^{(D)}= \frac{3}{4\cosh^4 r} -
 \frac{1}{\cosh^2 r}\,
 \left [
 N(N+1) + ( 2N+1) \delta_N + 1 + \frac{\beta^2}{\delta_N^2}
 \right ] +
 \een
 \be
+ \left ( \frac{\beta^2}{\delta_N^2} - \delta_N^2 \right )
 + 2i\beta\frac{\sinh r}{\cosh r}, \ \ \ \ \ \ \
\delta_N=A-N-1  >0.
 \label{exwpot}
  \ee
In the domain of the large $|r| \gg 1$ this formula is dominated
by the last two terms. Only the very last one depends on the sign
of ${\rm Re}\ r$ so that the coupling $\beta$ must be independent
of $N$ (we do not wish to have a state-dependent potential). We
then determine (i.e., strictly speaking, remove the
shift-ambiguity of) the energy $E^{(D)}$ by the convenient
requirement that $V^{(D)}(\pm \infty)$ vanishes at $\beta=0$. The
other two asymptotically smaller components of $V^{(D)}(\xi)$ are
of the first and second order in $1/\cosh^2r(\xi)$. Both of them
must be also independent of $N$ of course. In the first order this
gives the strict rule
 \be
 N^2+N+1 + (2N+1)\,\delta
 +\beta^2/\delta^2 = constant \ (= C).
 \label{map}
 \ee
In the second order, the coefficient is equal to $3/4$ and the
condition remains trivial. We just confirmed that the replacement
of $r$ by $r(\xi)$ transforms the Eckart potential (\ref{Eckartp})
into its exactly solvable descendant (\ref{hulth}), indeed.

During the re-construction of the potential $V^{(D)}(\xi)$,
energies $E^{(D)}_N$ and wave functions $\psi^{(D)}_N(\xi)$ the
auxiliary, $N-$dependent value of $\delta = \delta_N= A-N-1>0$ is
to be determined as a root of the cubic equation (\ref{map}). In
Hermitian setting, the correct account of the physical boundary
conditions makes this root {unique} \cite{newRR}. In the
generalized, ${\cal PT}$ symmetric setting, the exceptional
boundary condition in the origin becomes {redundant}. This is the
reason why we have chosen our particular example. {\it A priori},
one may expect that the choice of the root $\delta_N$ could be
{ambiguous}.

In the light of our present construction, the latter expectation
proves fulfilled. At the sufficiently large values of $C\geq
C_{min}>0$ there exist three real roots $\delta_N$. Only one of
them (viz., the negative one) can be eliminated as violating the
asymptotic physical boundary conditions (i.e., the normalizability
of the wave function). In contrast to the Hermitian case, two of
the roots $\delta=\delta^{(\pm)}= \delta^{(\pm)}_N(\beta,C)>0$ of
our cubic eq. (\ref{map}) remain equally acceptable. At any number
of nodal zeros $N$, each of them defines a separate energy level,
 \be
 E=E^{(D)(\pm)}_{N}=
 \left (
 \delta^{(\pm)}_N+N+\frac{1}{2}
 \right )^2 + \frac{3}{4}-C.
 \label{energies}
 \ee
The related ${\cal PT}$ symmetric wave functions
  \ben
 \psi=\psi^{(D)}_N(\xi)
 = \sqrt{\xi'[r(\xi)]}\cdot \psi^{(E)}_N[r(\xi)]
  \een
are proportional to the same Jacobi polynomials as above,
 \ben
\psi =  e^{-i \delta\,\xi}\,
 \left[
1-e^{-2i\xi} \right ] ^{1/2}
 \,
 \left [
\sqrt{ e^{2i\xi}-1}-e^{i\xi}
 \right ]^{i\beta/\delta}\,
 P_N^{( [ \delta-i\beta/\delta ] /4,
 [ \delta+i\beta/\delta ]/4)}
 \left ( \sqrt{1-e^{-2i\xi}}
 \right ) .
 \een
This is the core of our present message. In the $|t| \gg 1$
asymptotic regions we have $\xi \approx -iZ\pm \pi/2$ with $Z \gg
1$. This re-confirms that both our series of wave functions
$\psi^{(\pm)}$ are asymptotically vanishing as $\exp
(-\delta^{(\pm)} Z)$. For both the roots $\delta =
\delta^{(\pm)}_N>0$, they are safely normalizable.


\section{Discussion}
\label{cyzih}

The use of the commutativity $[H, {\cal PT}]=0$ complies with our
intuitive expectations in the numerical setting \cite{nume} and in
perturbation theory \cite{Caliceti}, in the WKB approximation
\cite{BB} and in the supersymmetric context \cite{sCann} as well
as in the phenomenologically oriented field-theoretical studies
\cite{BM}. In contrast, our present results form a paradox since
the change of variables mediates an utterly unusual {\em
one-to-two} correspondence between the two complex potentials. For
the force $V^{(D)}(\xi)$ this gives the two parallel series of
bound states which must be distinguished by an additional,
parity-type quantum number $q = \pm 1$. The nodal count $N$ itself
does not suffice to characterize the energy levels
(\ref{energies}). This is a puzzling situation since we see no
obvious reason for the introduction of $q$. One has to assume the
existence of some unknown, hidden symmetry in our problem
(\ref{SEoxi}), but it is still necessary to accept the fact that
this symmetry has to break down during an apparently innocent
change~(\ref{tren}) of the coordinates.

We can summarize that our above conclusions are to be added to the
list of the ``oddities" which emerge in ${\cal PT}$ symmetric
quantum mechanics due to it weaker boundary conditions. Let us
just remind the reader that this is not in fact an isolated
paradox. Our traditional intuition has already had really hard
times with the unavoided level crossings in ref. \cite{ptho}, with
the decrease of the energy with $N$ in ref. \cite{Morse}, with the
high excitations caused by a weak potential in refs. \cite{shape}
etc. Thus, just another item is provided by our present example.

We may formulate our tentative conclusion that the ${\cal PT}$
symmetric deformations of the integration paths can in fact
destroy the ({\it a priori}, plausible) similarity between the
complex and ordinary parity. One of the most important
implications is that any future appropriate generalization of the
Hermitian Sturm--Liouville oscillation theorems will be
necessarily not entirely trivial. Also the closely related concept
of the completeness of states~\cite{genera} must be dealt with an
exceptional care in any future development of the ${\cal PT}$
symmetric quantum mechanics. We might point out that the similar
words of warning have been recently issued also on the purely
numerical basis~\cite{BBSL}.

\section*{Acknowledgment}

This research was partly carried out within the frame of TMR -
Network ERB FMR XCT 96-0057, and partially supported by the grant
No. A 1048004 of GA AS CR (Czech Republic) and OTKA grant No.
T031945 (Hungary).

\end{document}